\documentclass[11pt,a4paper]{article}
\usepackage{bm,amssymb}

\newcommand\ie{\textit{i.e.}}
\newcommand\eg{\textit{e.g.}}
\newcommand\cf{\textit{cf.}}

\newcommand{\half}{\frac{1}{2}}

\newcommand{\be}{\begin{equation}}
\newcommand{\ee}{\end{equation}}
\newcommand{\bea}{\begin{eqnarray}}
\newcommand{\eea}{\end{eqnarray}}
\newcommand{\eqref}[1]{(\ref{#1})}

\newcommand{\dt}{\delta t}
\newcommand{\Thet}{\theta_{23}}
\newcommand{\phio}{\nu_\mu}
\newcommand{\phit}{\nu_\tau}
\newcommand{\Dm}{\Delta m^2_{23}}
\newcommand{\Dmm}{|\Delta m^2_{23}|}
\newcommand{\Em}{E_{med}}

\begin{document}
\begin{titlepage}
\begin{flushright}
{\tt SHEP 11-30}
\end{flushright}

\begin{center}
{\huge \bf  Superluminal group velocity through near-maximal neutrino oscillations}
\end{center}
\vskip1cm

\begin{center}
{\bf Tim R. Morris}
\end{center}

\begin{center}
{\it School of Physics and Astronomy,  University of Southampton\\
Highfield, Southampton, SO17 1BJ, U.K.}\\
\vspace*{0.3cm}
{\tt  T.R.Morris@soton.ac.uk}
\end{center}

\abstract{Recently it was suggested that the observation of
superluminal neutrinos by the OPERA collaboration may be due to
group velocity effects resulting from close-to-maximal oscillation
between neutrino mass eigenstates, in analogy to known effects in
optics. We show that superluminal propagation does occur through
this effect for a series of very narrow energy ranges, but this phenomenon
cannot explain the OPERA measurement.}

\end{titlepage}

\section{Introduction}

Recently the OPERA collaboration reported a measurement
of the average time taken for neutrinos ($\nu_\mu$ up to \% level
contamination) created at CERN (CN) to arrive at the Gran Sasso
Laboratory (GS) compared to the time taken travelling at the
speed of light in vacuo ($c$). They found an early arrival time
of approximately $\dt= 60 $ ns, which corresponds, at a
significance of $5.0\sigma$, to faster-than-light travel by a positive fraction $\delta v = v-1\approx 2.37 \times 10^{-5}$ \cite{OPERA}. (In this paper we set Planck's constant and the speed of light in vacuo $\hbar=c=1$.)

The OPERA result has inspired 
many papers (for a selection see \cite{some,Mecozzi,Tanimura,Berry,ED}).
Here we focus on the one by authors  Mecozzi and Bellini \cite{Mecozzi}. They have suggested an interesting possible interpretation in one-to-one analogy with the established dispersion and interference effects found in Mach-Zehnder interferometers and optical fibres with polarisation mode dispersion, where a superluminal group velocity has been both predicted and measured \cite{Wang}--\cite{Gisin}. 

This effect does not imply that a signal is exchanged at faster than the speed of light, in violation with special relativity. Instead it arises from constructive and destructive interference deforming the leading and trailing edges of the pulse. 

Mecozzi and Bellini do not analyse whether such superluminal propagation is possible in practice in OPERA, and indeed in neutrino experiments in general. In this paper we fill that gap. We show that remarkably such an effect does occur in the OPERA
experiment, causing superluminal `spikes' in the neutrino velocity at various `critical' values of the neutrino energy inside the neutrino beam energy spectrum. However this behaviour results in the wrong energy dependence and, once averaged over the full energy spectrum, is far too small to explain the OPERA result. 

Even though the spikes of superluminal group velocity caused by near-maximal neutrino oscillations cannot explain the OPERA measurements,  it is an effect that is nevertheless interesting in its own right. We analyse carefully the size of the effect and the conditions that are required to realise it. In particular, we point out that the value of the maximum achievable superluminal velocity depends crucially on corrections which were neglected in ref. \cite{Mecozzi}. We will see that these corrections recover the result one would expect intuitively: neutrinos do not escape the domains of their mass-eigenstate wave packets, which travel at the expected subluminal speeds, therefore the maximum increase in speed is entirely due to the quantum mechanical uncertainty in position divided by the time of flight.


We note that this effect has by now been treated in two other papers. In ref \cite{Tanimura}, it was pointed out that the OPERA experiment might be understood as a manifestation of weak measurement. This is in fact another way of interpreting the same effect however, as in ref. \cite{Mecozzi}, there is no attempt to determine whether this is realised for neutrinos in practice. 

Independently, the weak measurement interpretation was also treated in a paper that appeared after ours \cite{Berry}, where also a numerical estimate is provided. This is in agreement with our more detailed calculations, however the limiting corrections, multiple peaks and energy spectrum are not treated. On the other hand the treatment of the effect is performed in position space with Gaussian wave packets, which  exposes particularly clearly the root cause as being through the small relative displacement of the mass-eigenstate wave packets and their near-cancellation. 

We will see that in order to maximise the effect we will need a mixing angle such that $\sin^2(2\theta)$ is as close to 1 as possible. It is known that $\theta_{12}\approx 0.86$, however the effective value of $\sin^2(2\theta_{12})$ in rock is close to zero \cite{PDGrev}. $\theta_{13}$ is tiny \cite{recent13}, so we are therefore left with $\theta_{23}$ for which only the limits $0.92\lesssim\sin^2(2\theta_{23}) \le 1$
are so far known \cite{PDGrev}. $\theta_{23}$ drives $\nu_\mu\leftrightarrow\nu_\tau$ mixing and takes place practically as in vacuum \cite{PDGrev}. Clearly then we can work effectively with 2-neutrino mixing between $\nu_\mu$ and $\nu_\tau$. (Actually, this is not quite true: we know that the other mixing matrix parameters will supply corrections in exceptional circumstances. This will be discussed in the conclusions.)

The rest of the paper is organised as follows. In the next section
we review the derivation in ref. \cite{Mecozzi}, including the
notation and adaptations we will need to match the OPERA
experiment. In particular we take care to include corrections neglected in ref. \cite{Mecozzi} but which are crucial in limiting the maximum size of the effect. To carry through the derivation, we will need to make
some assumptions about the coherence of individual neutrino wave
packets. In section 3 we show these are correct. In section 4 we
show that while there is an effect, it is far too small and has the
wrong energy dependence to agree with the data.

In section 5, we draw our conclusions. 

\section{Review and development of ref. \cite{Mecozzi}}

Our treatment follows closely ref. \cite{Mecozzi} but with adaptations and corrections. We write the mass eigenstates in standard convention \cite{PDGrev} as:
\bea\label{estates} |p, \nu_2\rangle &=& \cos(\Thet) |p, \phio\rangle - \sin(\Thet) |p, \phit\rangle, \\
|p, \nu_3\rangle &=& \sin(\Thet) |p, \phio\rangle + \cos(\Thet) |p, \phit\rangle, \eea
where $p$ is the momentum of the neutrino in the beam. For momenta much larger than the masses, which is the case here (since the masses are known from tritium decay to be $\lesssim2$ eV \cite{PDGrev}), energies are given by
\be E_2 = E_0 + E/2\quad\textrm{and}\quad E_3= E_0-E/2\ee
where
\be\label{Es} E_0(p) = p+{m^2_2+m^2_3\over4p}\equiv p +{\langle m^2_{23}\rangle\over 2p}\quad\textrm{and}\quad E(p) = {{m_2^2-m_3^2}\over2p} \equiv {\Delta m^2_{23}\over2p}.\ee

We start at CERN with a normalised pure $\nu_\mu$ wave function $|\psi_0\rangle$ at time $t=0$, and position $x_0 = \langle\psi_0| x|\psi_0\rangle$. Projecting on mass eigenstates using $\sum_i|p',\nu_i\rangle\langle p',\nu_i|$, we have at later times
\be |\psi_t\rangle = \int \!\!d k \, \langle k |\psi_0\rangle\, \textrm{e}^{-iE_0 t}
 \big\{\sin(\Thet) \,\textrm{e}^{-i E t/2} |k, \nu_3\rangle  +\cos(\Thet) \,\textrm{e}^{i E t/2} |k, \nu_2\rangle \big\},\ee
where $\psi_0(k)=\langle k |\psi_0\rangle$ describes the spread of momenta in the initial neutrino wave packet.

To find the $\nu_\mu$ wave function when measured at Gran Sasso, we collapse $|\psi_t\rangle$ using $\langle p',\nu_\mu|$ to obtain
\be
\label{GS}
\psi'(p') =
\psi_1(p') F(p')
\ee
where
\be\label{F}
 F(p') = \left\{ \,\textrm{e}^{-i E t/2}\sin^2\Thet +\,\textrm{e}^{i E t/2}\cos^2\Thet\right\}\, \textrm{e}^{-i E_0 t},
\ee
and we have replaced the initial $\psi_0(p')$ with an effective measured $\psi_1(p')$ to allow for further decoherence in the momentum space of the particle during its measurement.

We need to normalise \eqref{GS}, which requires dividing by $ \int\!\!dp' |\psi'(p')|^2$. We now assume, as effectively was done in ref. \cite{Mecozzi}, that $\psi_1(p')$ is strongly peaked around the momentum $p'=p$, and that the variation in $|F(p')|$ is gradual in comparison. We will confirm these assumptions in the next section. 
We note that in this case $ \int\!\!dp' |\psi'(p')|^2\approx |F(p)|^2$ and so the Gran Sasso wave function is typically normalised to good approximation by dividing by $|F(p)|$. This approximation breaks down if $p$ takes a value such that $F(p)$ is close to vanishing (as is clear because $ \int\!\!dp' |\psi'(p')|^2$ is positive definite since $F$ cannot vanish for all momenta). We will return to this in the next section.

For now we take the expected position of the $\nu_\mu$ at Gran Sasso, to thus be given by
\be
\label{position}
\langle x\rangle \approx {1\over |F(p)|^2}\int\!\! dp' \psi'^*(p')\, i{\partial\over\partial p'}\, \psi'(p'),
\ee
which depends on the time of arrival through $F$. Substituting \eqref{GS}, we have that the integral is given by $N_1+N_2$ where
\be 
N_1 = \int\!\!dp'\,|F|^2\psi_1^*\, i{\partial\over\partial p'}\, \psi_1\quad\textrm{and}\quad N_2 = \int\!\!dp'\,|\psi_1|^2 F^*\, i{\partial\over\partial p'}\, F.
\ee
Using again the fact that $\psi_1$ is strongly peaked we see that
\be
N_2 \approx F^*(p)\,  i{\partial\over\partial p}\, F(p).
\ee
Since \eqref{position} is the expectation of an Hermitian operator it must give a real answer. Therefore the imaginary part of the above must get cancelled. This can be seen to be true by integrating $N_1$ by parts to get:
\be
N^*_1 = \int\!\!dp'\,|F|^2\psi_1^*\, i{\partial\over\partial p'}\, \psi_1 +  \int\!\!dp'\,|\psi_1|^2 \, i{\partial\over\partial p'}\, |F|^2.
\ee
and thus extract the imaginary part. The remaining real part of $N_1$ is approximately
\be
 |F|^2(p)\int\!\!dp'\,\psi_1^*\, i{\partial\over\partial p'}\, \psi_1\approx
 |F|^2(p)\, x_0.
 \ee
Putting it all together we see that
\be
\langle x\rangle = x_0+v_g t \approx x_0 + {1\over |F(p)|^2}\Re \left\{F^*(p) \, i{\partial\over\partial p}\, F(p)\right\},
\ee
where we have identified the second term as $t$ times the group velocity. Finally, substituting \eqref{F} and using \eqref{Es}, we have
\be
\label{modFp}
|F(p)|^2 = 1-\sin^22\Thet\,\sin^2({\Dmm t/(4p)})
\ee
and evaluating the numerator:
\be
\label{vg}
v_g = 1 + \delta v_{cl} +\delta v_{sup}
\ee
where
\be
\label{vcl}
\delta v_{cl} = - {\langle m^2_{23}\rangle\over2p^2}
\ee
is the expected negative `classical' correction to the speed of light as a consequence of the average squared mass of the two mass eigenstates,  
and the last term is a correction resulting from interference between the two mass eigenstates:
\be
\label{vsup}
\delta v_{sup}\approx {\Dm\over4p^2}{\cos2\Thet\over1-\sin^22\Thet\,\sin^2({\Dmm t/(4p)})}.
\ee
This is the main result of ref. \cite{Mecozzi} (up to correction of the power of $\sin2\Thet$ and the explicit statement of approximation).

\section{Limitations to coherence and the strongly-peaked wave packet approximation}

Let us revisit the assumption that the momentum of the measured neutrino wave packet $\psi_1(p')$ is strongly peaked about $p'\approx p$. This was used to derive the key formulae in sec. 2 below \eqref{F}, by assuming that the variation in $|F(p')|$ was by comparison more gradual. It should be clear that here we are not discussing the momentum spectrum of the ensemble of neutrinos in the beam, which is very broad ---  as we will describe in the next section, but rather the inherent quantum mechanical uncertainty in the momentum of a given neutrino as it is measured at Gran Sasso.

We start with the neutrino wave function $\psi_0(p')$ produced in the CNGS decay tunnel at CERN. The momentum uncertainty can certainly can be no smaller than that set by $\Delta t \approx 5$ ns, the smallest time features in the proton bunch \cite{OPERA}.\footnote{In the new version of the OPERA experiment $\Delta t \approx 3$ ns, as set by the width of the bunch.}This corresponds to $c\Delta p=1/\Delta t\approx1.3\times10^{-7}$ eV.

However, even if the proton beam is coherent at this level, it suffers decoherence on its way to becoming the $\nu_\mu$ beam. Firstly, the protons impact the graphite target, producing the mesons (mostly pions) that will decay to muon neutrinos. Initially these mesons are in a quantum state together with the other products of the collision (including various nuclei), however they then suffer decoherence from thermalisation in a hot target, both directly and also through their quantum mechanical coupling to the decay products. Assuming a target temperature of, say 300$^\circ$C,\footnote{This temperature is confirmed as the average maximum temperature in the revised version of the OPERA experiment \cite{OPERA}.} this limits the energy-momentum resolution to $k_BT\sim0.05$ eV.

Finally the mesons decay in the decay tunnel and here further decoherence takes place, again through coupling to the decay products (in this case the muon). Consider for example the decay of a $\pi^+$ to $\mu^+\nu_\mu$. The resulting quantum state takes the form:
\be
\label{pidecay}
\psi_\pi({\bf q})\int_{phase\ space}\!\!\!\!\!\!\!\! \mathcal{M}({\bf p},{\bf q})\  |\nu_\mu({\bf p})\rangle\,|\mu^+({\bf q}-{\bf p})\rangle,
\ee
where $\psi_\pi({\bf q})$ is the wave function of the erstwhile pion,  ${\bf q}$ and ${\bf p}$ are 3-momenta, and $\mathcal{M}$ stands for the matrix element for the decay. The muon is absorbed by a combination of rock, a Hadron stop and two muon detectors \cite{OPERA,CNGStalkonwebsite}. This allows the experimenters to measure the transverse coordinates of the proton beam spot when it hits the target to a precision of $\sim$ 50 -- 90 $\mu$m \cite{CNGStalkonwebsite}, however it is reasonable to assume that the rock itself localises the muons at the $\mu$m level (similar to emulsion --- see below), even if this is not recorded. This corresponds to a momentum decoherence of order 0.2 eV/c. Through \eqref{pidecay} this decoherence is transferred to the neutrino.

We conclude that the chief limiting factor on the coherence of the initial neutrino wave packet is through the decay of the mesons and results in a wave packet with momentum uncertainty $w\sim \pm0.1$ eV/c.

For muon neutrinos that interact in the Gran Sasso detector, the impact spot is discernible in the emulsion at the $\mu$m level \cite{impactspot}; we can expect similar localisation in the rock in front of the detector for the external events. Therefore the act of measurement results in a momentum spread in the wave packet of similar size to that in the initial packet.

We now contrast this with the variation in $|F(p')|$ from
\eqref{F}. We see that this is controlled by the phase
\be
\phi(p')=|Et| \equiv  {\Dmm t\over2p'},
\ee
using \eqref{Es}, where $t$ can be taken to be $L/c$ and
\be
\label{L}
L = 730,085\, \textrm{m}
\ee
is the corrected distance from the average meson decay point in the CNGS decay tunnel to the origin of the OPERA detector \cite{OPERA}. 
At the average energy $p=17$ GeV/c of the neutrino beam \cite{OPERA}, and using $\Dm\approx2.43\times10^{-3}$ eV$^2$ \cite{PDGrev}, we have that $\phi$ has magnitude 0.26, and thus we see that $|F(p')|$ is indeed slowly varying even on changes of order $\Delta p' \sim$ GeV. 

More generally we require 
\be
\label{condition}
w\left|{\partial\phi\over\partial p'}\right| = {1\over2}\left({30.0\, \textrm{keV}\over p'}\right)^2\ll 1.
\ee
The minimum energy-momentum $p'$ we can consider is set by the resolution of the Gran Sasso detector, which is $E_{min}\approx1$ GeV \cite{impactspot}, or more optimistically by the lowest threshold energy for the charge current interactions used to detect the $\nu_\mu$. This occurs for the process $\nu_\mu n\to \mu^- p$, which would imply $E_{min} > m_\mu+{\half m_\mu^2/m_n}=112$ MeV. Either way we see that the condition \eqref{condition} is amply satisfied.

Although this establishes that $|F(p)|$ is  slowly varying compared to the fundamental uncertainty $w$ set by the width of the wave packet, as we mentioned in sec. 2 we cannot trust \eqref{vsup} if $F(p)$ is close to vanishing. This leads to extra conditions that must be satisfied if we are to trust the approximation. From \eqref{modFp}, we see that $|F(p)|$ takes its minimum value, $\cos^22\Thet$, when 
$\phi(p)=\pi+2\pi a$, where $a=0,1,2,\cdots$.  This corresponds to `critical' values of momentum
\be
\label{crit}
p=p_{crit}(a) = {\Dmm t\over2\pi}{1\over1+2a} = {1.43\over1+2a}\, \textrm{GeV/c}.
\ee
Taylor expanding about these values we have
\be
|F(p')|^2 = \cos^22\Thet+{\pi^2\over4p^2_{crit}}(1+2a)^2\sin^22\Thet[\Delta p']^2+O([\Delta p']^4),
\ee
where $\Delta p' = p'-p_{crit}$. Thus from \eqref{GS}, at $p=p_{crit}$, we can compute the first correction to $ \int\!\!dp' |\psi'(p')|^2\approx |F(p)|^2$, from the finite size of the wave packet. It is independent of the details other than through the RMS width $w^2 = \langle \psi_1| [\Delta p']^2|\psi_1\rangle$: 
\be
\label{corrections}
 \int\!\!dp' |\psi'(p')|^2 = \cos^22\Thet+{\pi^2\over4}{w^2\over p^2_{crit}}(1+2a)^2\sin^22\Thet+O(w^4/p^4_{crit}),
\ee
Thus if 
\be
\label{c1}
|\cot2\Thet|\gg{\pi\over2}{w\over p_{crit}}(1+2a) 
\ee
the first term in \eqref{corrections} dominates and $ \int\!\!dp' |\psi'(p')|^2\approx |F(p)|^2$ is always a good approximation. If \eqref{c1} is not satisfied then $p$ cannot be taken too close to $p=p_{crit}(a)$: we can only trust the expression \eqref{vsup} for $\delta v_{sup}$ providing
\be
\label{c2}
|F(p)|^2\gg \cos^22\Thet+{\pi^2\over4}{w^2\over p^2_{crit}}(1+2a)^2\sin^22\Thet.
\ee

\section{Confronting the experiment}

Given the average  17 GeV energy of the beam, the classical
correction to the speed of light \eqref{vcl} is negligible and can
be neglected. For most values of $p$,  $\delta v_{sup}$ is also clearly negligible.

However, \eqref{vsup} provides a correction that is maximal at the critical values of momenta \eqref{crit}.
Note that the requirement of detection at Gran Sasso sets the upper limit to $a$ as 
\be
a< {\Dmm t\over4\pi E_{min}}-{1\over2}.
\ee
This implies $a$ can only be zero if $E_{min}=1$ GeV, or $a\le6$  for $E_{min}=112$ MeV.

If $\theta_{23}$ is infinitesimally close to the maximal mixing value of $\pi/4$, then \eqref{vsup} actually diverges at the points $p=p_{crit}(a)$.  To see this more clearly, let
\be
\label{smallshifts}
2\Thet = {\pi\over2}\mp\delta\theta\quad\textrm{and}\quad p=p_{crit}+\epsilon,
\ee
where the sign is opposite to that of $\Dm$, $\delta\theta>0$ is a small shift from maximal mixing, and $\epsilon$ is a small deviation in energy from the critical value. Then \eqref{vsup} becomes
\be
\label{vgd}
\delta v_{sup} \approx {\Dmm\over4p_{crit}^2}{\delta\theta\over(\delta\theta)^2+\pi^2\epsilon^2/(4 p_{crit}^2)}.
\ee
However, at $\epsilon=0$, from \eqref{c1}, this is only trustworthy if 
\be
\label{trusty}
|\delta\theta|\gg {\pi\over2}{w\over p_{crit}}(1+2a) = 1.10\,(1+2a)^2\times10^{-10}
\ee
Substituting this in \eqref{vgd} and using \eqref{crit}, we see that the maximum superluminal effect that can be predicted is
\be
\label{qmbound}
\delta v_{sup} \ll {1\over wt} = 2.70\times10^{-12}.
\ee
This formula has a simple intuitive explanation. $1/w$ is the spatial width of the neutrino wave packet. Therefore the upper limit on $\delta v_{sup}$ is nothing but the maximum increase in speed that could result from the quantum mechanical uncertainty in the position of the neutrino. We see that in order to explain the OPERA measurement we would have to assume that the width of the neutrino wave packet created at CERN is  $1/w\gg \delta t=$ 60 ns which is not compatible with the analysis in the previous section.

We note also that OPERA repeated the analysis concentrating on only those $\nu_\mu$ charged current events occurring in the OPERA target (where reliable energies could be measured). They split this sample into bins of nearly equal statistics, taking events with energy higher or lower than $\Em=20$ GeV. With a significance of greater than 3$\sigma$, they still see a superluminal velocity in the higher energy sample of the same magnitude (within errors) \cite{OPERA}. If this group velocity effect was the explanation then we ought to find that the effect goes away in the higher energy bin.

Finally let us show that, taking into account the spread in energies in the neutrino beam, the superluminal contribution from this effect is in fact some 12 orders of magnitude smaller than \eqref{qmbound}.  Only a very narrow spread in energies in the neutrino beam contributes significantly to \eqref{vgd}. We see that the energies contributing must lie in the range
\be
\label{range}
p=p_{crit}\pm\epsilon,\quad \textrm{with}\quad\epsilon\lesssim 2p_{crit}\delta\theta/\pi 
\ee

In fact, from the talk given at CERN \cite{talk} and earlier OPERA analysis \cite{OPERAMC}, we can see that the energy spectrum is broad, rising approximately linearly from $E_{min}$ to a maximum at $E= \Em$, and then falling with a large tail reaching energies of $\sim$ 200 GeV. Let the fraction of neutrino events measured between momenta $p$ and $p+dp$ be $n(p)$. If we model $n(p)=\rho p$ for $p< \Em$, where $\rho$ is a constant, then, since about half the events are found below $\Em$, we have by $\rho \approx 1/\Em^2$. It follows that $n(p_{crit})\approx p_{crit}/\Em^2 = 3.6\,(1+2a)^{-1}\!\times10^{-12}$ eV$^{-1}$. Since \eqref{vsup} is sharply peaked around $p=p_{crit}$, the average contribution from \eqref{vgd} is to very good approximation
\be
\label{avsup}
\int\!\! dp\, n(p)\, v_{sup}(p) = {1\over2}\Dmm \sum_a{n(p_{crit})\over p_{crit}} \approx {A\over2}{\Dmm\over \Em^2},
\ee
where $A$ is the number of peaks (1 or 7 depending on $E_{min}$).
This is numerically $3A\times 10^{-24}$, which is of course too small to measure. It is also competitive with the classical term \eqref{vcl}. Using the same model for $n(p)$ we have for the classical term
\be
\label{avcl}
\int\!\! dp\, n(p)\, v_{cl}(p) < - {1\over2}{\langle m^2_{23}\rangle\over E_{med}^2}\ln\left({E_{med}\over E_{min}}\right),
\ee
where the inequality results from neglecting the contribution of energies greater than $E_{med}$. This term will dominate unless there is a large hierarchy between the two masses, in which case this is of the same order, making the detected average group velocity slower (faster) than the speed of light, depending on whether the larger (smaller) value of $E_{min}$ is used.

\section{Conclusions}

We have seen that if the mixing angle $\Thet$ is close to maximal then thanks to intereference between the two mass eigenstates, the group velocity for the muon neutrino wave packet, \eqref{vg} using \eqref{vcl} and \eqref{vsup}, develops a series of very sharp peaks at critical momenta
\be
\label{peaks}
p=p_{crit}(a) = {\Dmm t\over2\pi}{1\over1+2a},
\ee
where $a$ is a non-negative integer. It is surely interesting in its own right that superluminal propagation of neutrinos is possible in principle through such an effect. In the OPERA experiment this corresponds to energies $1.43/(1+2a)$~GeV and in practice $a$ is bounded above by the fact that for sufficiently large $a$ it is no longer possible to detect the neutrino, definitely $a\le6$. 

However from \eqref{qmbound} we have seen that the maximum extra displacement caused by this is limited above by $\Delta x=1/w$, where $\Delta x$ is the width of the wave packet and $w$ is the width in momentum space. This translates into a maximum increase in group velocity of $\Delta x/t$, where $t$ is the na\"ively expected time of flight. Therefore we see that the effect is not in conflict with relativity and causality, being limited to no more than that caused by the inherent quantum mechanical uncertainty in position. We have argued that the wave packet has a width of order 1 $\mu$m, as a result of decoherence from the detection (whether recorded or not) of the associated muons. Even if this is not the case we noted that thermalisation in a hot target would limit the coherence length of the pion wave packet, and thus also the neutrino wave packet to $\sim 4 \mu$m. This corresponds to a maximum superluminal correction of $\delta v_{sup} \approx 10^{-11} c$ and is  therefore far too small to correspond to the effect seen by OPERA.

We note that this also provides potentially a severe constraint on proposals to explain the OPERA measurement via superluminal propagation of only certain mass eigenstates (\eg\, a sterile neutrino in extra dimensions \cite{ED}) since the OPERA measurement corresponds to an extra displacement by approximately 20 m, whereas a relative shift between mass eigenstate wave packets of more than a few $\mu$m is enough to destroy the coherence of the observed neutrino oscillations. 

We do not reach the upper bound, $\Delta x/t$, on the maximum extra group velocity unless the mixing angle  $\theta_{23}$ is closer to maximal than a term of order $w/p$, \cf\,  eqn. \eqref{trusty}. Such finite-width corrections are negligible everywhere except in this case, and then only when close to the group velocity peaks, where they matter because of they correct the near vanishing of the denominator term in \eqref{vsup}. There are some corrections to the numerator that become important similarly since it is also close to vanishing, although we did not need them for our analysis. In practice the neglected small corrections due to the other neutrino mixing parameters ($\delta$, $\theta_{13}$ and the effective value of $\theta_{12}$ in rock) could also have an effect when close to these peaks for the same reason. The exact behaviour close to the peaks is therefore the result of a  competition between these corrections, and will depend on which one dominates.

None of this matters for the measurements performed in the OPERA experiment however since they measure an effect that is averaged over the broad energy spectrum of the neutrinos. 
The average effect would give a strong dependence on energy, which is not seen in the OPERA experiment, if it weren't for the fact that the result is anyway too small to measure. Furthermore we have seen that, once averaged over the energy spectrum of the neutrino beam, the resulting net correction \eqref{avsup} is $\sim 10^{-23}$, at best of the same order as the negative (\ie\, subluminal) classical correction.

It seems very difficult to imagine a scenario where this effect could explain the OPERA measurement. We note that the neutrinos with superluminal group velocity would be found preferentially at the leading edge of the neutrino bunch. But from the plots \cite{OPERA,talk}, it is clear that there is no tail of early arrivers: the shift due to an excess of superluminal neutrinos at the leading edge cannot be much more than the measured $\delta v$, and besides we have too few of them by this effect in the energy spectrum. Similar comments apply to the superluminal-depleted population at the trailing edge. With the new measurements using shorter pulses \cite{OPERA} all such types of explanation are ruled out.

\section*{Acknowledgments}
The author  thanks the STFC for financial support.

\end{document}